\documentclass[sigconf]{acmart}

\setcopyright{none}

\usepackage{graphicx} 
\usepackage{hyperref}
\usepackage{tikz}
\usepackage{pgfplots}
\usepackage{pgfplotstable}
\pgfplotsset{compat=1.7}
\usepackage{subcaption}
\usepgfplotslibrary{groupplots}


\copyrightyear{2024}
\acmYear{2024}
\setcopyright{acmlicensed}\acmConference[IDE '24]{2024 First IDE
Workshop}{April 20, 2024}{Lisbon, Portugal}
\acmBooktitle{2024 First IDE Workshop (IDE '24), April 20, 2024, Lisbon,
Portugal}
\acmDOI{10.1145/3643796.3648460}
\acmISBN{979-8-4007-0580-9/24/04}

\begin{document}

\title{Trigram-Based Persistent IDE Indices with Quick Startup}

\author{Zakhar Iakovlev}
\affiliation{
    \institution{ITMO University}
    \country{Russia}
}

\author{Alexey Chulkov}
\affiliation{
    \institution{ITMO University}
    \country{Russia}
}

\author{Nikita Golikov}
\affiliation{
    \institution{ITMO University}
    \country{Russia}
}

\author{Vyacheslav Lukianov}
\affiliation{
    \institution{Huawei RRI}
    \country{Russia}
}

\author{Nikita Zinoviev}
\affiliation{
    \institution{Huawei RRI}
    \country{Russia}
}

\author{Dmitry Ivanov}
\affiliation{
    \institution{Huawei RRI}
    \country{Russia}
}

\author{Vitaly Aksenov}
\affiliation{
    \institution{City, University of London}
    \country{UK}
}

\begin{abstract}
One common way to speed up the find operation within a set of text files involves a trigram index.
This structure is merely a map from a trigram (sequence consisting of three characters) to a set of files which contain it.
When searching for a pattern, potential file locations are identified by intersecting the sets related to the trigrams in the pattern.
Then, the search proceeds only in these files.

However, in a code repository, the trigram index evolves across different versions. Upon checking out a new version, this index is typically built from scratch, which is a time-consuming task, while we want our index to have almost zero-time startup.

Thus, we explore the persistent version of a trigram index for full-text and key word patterns search.
Our approach just uses the current version of the trigram index and applies only the changes between versions during checkout, significantly enhancing performance.
Furthermore, we extend our data structure to accommodate CamelHump search for class and function names.

\end{abstract}

\maketitle

\section{Introduction}

An Integrated Development Environment (IDE) is a pivotal tool for programmers, offering more than a mere text editor by incorporating numerous essential features. Among these, the text search stands out. The basic approach of searching through all project files using algorithms like the Z-function~\cite{gusfield1997algorithms} tends to be slow for modern projects with extensive repositories housing multiple versions. To ensure quicker response times, IDEs rely on index data structures. This approach can be used both for backend and cloud services, and for personal usage. However, we suggest to use our approach on remote services.

Various index implementations facilitate full-text searches~\cite{compressed-full, fulltext-genome}. One prominent approach is the utilization of a \emph{trigram index}~\cite{trigram} that is merely a map: from \emph{(char, char, char)} to the \emph{set of files} containing it. In contemporary IDEs like IntelliJ IDEA~\cite{intellij2011most} indices are limited to the current working copy, encompassing files in the current project revision alongside any uncommitted in-memory changes. The main goal of this work is to create an efficient approach that can maintain the current working index and can restore the index for any version.
The proposed trigram index aims to support several desired features in modern IDEs: 1)~zero-time startup, 2)~an enhanced code review, and 3)~streamlined navigation through history, discussed individually below.

\paragraph{Zero-time startup.}
If a developer wants to work with a specific git commit, current state-of-the-art indices has to be built for this commit from scratch, a process that could take several minutes or longer depending on the project size.
In contrast, our approach stores indices for all commits stored in a specific format, allowing us to build the index for any commit within seconds.

This feature is crucial for Cloud IDEs. They support just an editor for the client while the server maintains a git repository and corresponding indices.
Having this feature allows the developer to just click once on a chosen commit to obtain a fully-functional IDE.
Our goal is to make this as efficiently as possible, so that the change of the revision will be almost immediate for the user.

\paragraph{Enhanced code review.}
Cloud tools like GitHub/GitLab primarily present code review as a diff between two text files without robust navigation features. Our index structure can efficiently provide necessary navigation features for both versions in such cloud tools, enabling a more insightful code review.

\paragraph{Search and navigation through the history.}
Navigate and search features are extremely important in contemporary IDEs for the understanding of the code. For example, the ``Go-to-Class'' feature allows the user to navigate to any class by typing the substring from its name.
To simplify such search for the user IDEs support \emph{CamelHump} search~\cite{camelhumps}.
It helps to find specific code patterns, so-called \emph{symbols}, i.e., the names of classes, functions, fields, etc., not only by the full name, but also by its parts, when the user does not remember the exact name or does not want to write the whole name.
Our data structure supports this feature, enhancing navigation by allowing searches, for example, using the first letters of CamelCase words.

\paragraph{Related work.}
A trigram search~\cite{trigram} is a popular method for text analysis, widely used due to its compact index size.

Full-text search is often performed using string data structures, such as tries and suffix trees~\cite{compressed-full}. Although such data structures perform well when text is known in advance, and even under some moderate modifications, they fail to achieve our goal, as we need such data structure to be persistent.
However, traditional persistent data structures face limitations in large repositories as they fail to fit into RAM.
At the same time, hybrid RAM and on-disk approaches, e.g.,~\cite{utree}, do not satisfy us since they do not provide zero-time startup~--- one need to load the nodes in RAM first.
To achieve this requirement, our trigram index uses a key-value storage~\cite{keyvalue}: the current implementation uses LMDB~\cite{LMDB}.


\paragraph{Roadmap.}
In Section~\ref{sec:design} we explain the design of our approach. In Section~\ref{sec:experiments} we present the results of our experiments. Finally, we conclude in Section~\ref{sec:conclusion}.

\section{Design}
\label{sec:design}

The \emph{trigram} represents a sequence of three consecutive symbols within a text. These trigrams can help to find locations of a pattern: if we have the information for each trigram's locations in the text, we can intersect these locations to determine the pattern's positions. 
In addition, we can implement more advanced searches such as a CamelHump search by storing the information for specifically prepared trigrams.
Therefore, implementing an efficient map for trigrams is sufficient.

\subsection{Trigram-based persistent data structure for the full-text and CamelHump search}

Our primary aim is to create a persistent trigram index that facilitates two Git operations: 1)~checkout a specific version, and 2)~commit new changes. 
The checkout operation switches the current file version to a previously stored version in the repository, while the commit operation records changes made to the files in the current version.
%
It is important to note that the merge command in Git, which combines two different versions or branches, can be represented as a commit to one of the versions.

We maintain an on-disk trigram index using a key-value storage LMDB~\cite{LMDB}. For the ordered pairs of a trigram and a file we store the number of occurrences of this trigram in the file.
Our index supports various operations, including adding or removing files with a trigram, modifying trigram occurrences in a file, switching between revisions, and identifying all files with a specific trigram.

To support different revisions we store the data in the following way.
Initially, we build a \emph{global revision tree} where each vertex represents a revision and maintains a pointer to its parent~--- the previous version.
The structure of this tree resembles the Git tree without merges.
Moreover, each vertex holds the \emph{delta}, storing information necessary to manage the trigram index, such as a list of changes between that revision and its parent.
Hence, switching between revisions involves applying these deltas instead of traversing through all files.

\subsection{Operations}

We continuously maintain the trigram index for the current active revision including ongoing updates.

Now, we explain on how we implement our two main operations.

\begin{itemize}
    \item \textbf{The Checkout operation.}
    
    For performing a checkout to a specific version, we need to recalculate the trigram index and switch the active revision to the target version.

    The recalculation process follows these steps: suppose the current version is $v_1$, and the target version is $v_2$. The lowest common ancestor (LCA) of $v_1$ and $v_2$ in our global revision tree is denoted as $u$. To achieve our goal, we must invert all deltas along the tree path from $v_1$ to $u$ and subsequently apply all deltas along the tree path from $u$ to $v_2$.
    The invert and application of a delta affect the index which resides in the key-value storage.

    Therefore, such query is answered in $\mathcal{O}(\text{path size})$, where the path size refers to a total size of deltas along a path between $v_1$ and $v_2$ in the tree.
    Determining the LCA can be conducted online using any LCA algorithm~\cite{bender2000lca}.

    \item \textbf{The Commit operation.}
    
    Commit queries are straightforward: a new vertex is created and assigned the current revision as its parent. The maintained delta for the current version and the previous one (commit description) is dumped onto the disk and the active revision is changed to the created version. This operation works in $\mathcal{O}(\text{delta size})$.

    
\end{itemize}

\subsection{CamelHump search}

In general, the logic behind CamelHump search mirrors that of full-text search.
It leverages the trigram index, which stores specific trigrams.
%
%
However, it is important to note that the order and ranking of the results can vary depending on the implementation.

\subsubsection{CamelHump patterns}
Let us start by defining the patterns identifiable through CamelHump search. These patterns are substrings formed by merging the prefixes of consecutive CamelCase words.
For instance, from \texttt{``{CamelHumpSearch}''} we have the following valid patterns: \texttt{CHS}, \texttt{Camel}, \texttt{CamH}, and \texttt{CamHSearch}; while the following patterns are invalid: \texttt{CamelCase}, \texttt{CamelSearch}, \texttt{CS}, and \texttt{CelHump}.
Note that the CamelHump search is not case-sensitive, hence patterns like patterns \texttt{chs} or \texttt{cHS} are also valid.

These patterns allow for only specific character sequences after each letter: either the following (lowercase) letter within the current word or the first (uppercase) letter of the subsequent word. 
Thus, each word's initial letter can generate up to four trigrams.
For example, for the phrase \texttt{CamelHumpSearch} the trigrams starting from the letter \texttt{m} in \texttt{Camel} are \texttt{mel}, \texttt{meH}, \texttt{mHu} and \texttt{mHS}, which can be stored in lowercase depending on usage. The total number of stored trigrams does not exceed four times the combined length of all symbols. This combined size is much less than the total size of files. It makes storing the CamelHump data structure negligible in comparison to the full-text search one.

\subsubsection{CamelHump data structure}

Primarily, the trigram index for CamelHump search resembles that of normal search. However, for each trigram, instead of file appearances, we store symbol appearances such as class names and function names. 

The algorithm retrieves symbols for each trigram appearing in the search pattern and then intersects the results.
Note that the correctness of the found symbols should be checked, as trigrams can appear in different order and positions compared to the pattern.
Consequently, we need to calculate the ranking of relevance for the found symbols.



\subsubsection{CamelHump results ranking}

Before presenting the found symbols to the user, it is important to rank them. First, the pattern and the symbol must align: the pattern should represent a concatenation of hump prefixes. Key parameters for the matching include: 1)~the matching of the first letter of the pattern and the symbol; and 2)~the matching of the letter case of a hump starting with the corresponding pattern letter. The biggest penalty is given for skipped humps. The cases of other matching letters and the number of the humps are also important.

\section{Experiments}
\label{sec:experiments}

We implemented our approach and tested it on several open-source repositories. For the presentation, we chose three of the moderate size: \href{https://github.com/JetBrains/xodus/}{\textbf{xodus}}~\cite{xodus}, \href{https://github.com/apache/commons-lang}{\textbf{commons-lang}}~\cite{commons-lang}, and \href{https://github.com/apache/maven}{\textbf{maven}}~\cite{maven}. 

All experiments were conducted on Apple MacBook Pro with 8-core Apple M1 Pro CPU and 16 GB RAM and averaged over five executions. We made a preliminary integration in Visual Studio Code.

We implement both the standard trigram index and the CamelHump trigram index. For each repository, we parse \texttt{.git} files of the whole project, i.e., load all the revisions and build the global revision tree.
We measure the time and memory taken for the initialization of our data structure.
We also measure total memory stored on the disk by our approach (the size of the index and the total size of deltas combined).

As observed, the initialization process demands significant time and memory. Nonetheless, this initialization happens only once for the entire repository. 
After that, for a consistently growing repository, our program parses commits one at a time and works fast enough.
Indeed, for example, after the initialization for a repository with roughly $3\,000$ commits, the average processing time per commit is around $40$ms, which is sufficiently small.
Note, that we store the respective index on the disk, thus, we ensure the instantaneous startup for subsequent sessions, contributing to an almost zero-time startup.

\subsection{The trigram index}

\paragraph{Initialization}
We start with understanding how much the time is consumed during initialization, how much space all raw commits use, and the disk storage allocation for our global revision tree and the working index. The results are presented in Table~\ref{table:trigram}.

\begin{table}[ht]
\begin{tabular}{|l|c|c|c|c|}
\hline       & \begin{tabular}{@{}c@{}}
                Number \\ of \\ commits
             \end{tabular} & \begin{tabular}{@{}c@{}}
                Total \\ memory
                \\
                of commits, \\ MBs
             \end{tabular} & \begin{tabular}{@{}c@{}}
                Memory \\
                used, \\ MBs
             \end{tabular} & \begin{tabular}{@{}c@{}}
                Preprocess \\
                time, \\ sec            
                \end{tabular} \\ \hline
\textbf{xodus} & 3180 & 253 & 225 & 118 \\ \hline
\begin{tabular}{@{}l@{}}
\textbf{commons-}\\
\textbf{lang}
\end{tabular} & 7653 & 765 & 387 & 188 \\ \hline
\textbf{maven} & 14275 & 453 & 658 & 340 \\ \hline
\end{tabular}
\caption{The initialization of the trigram index. The number of commits and the total memory of commits are the characteristics of the repository. Memory used is the memory consumed by our data structure on the disk, and the preprocessing time is the time for the initialization of the whole repository.}
\label{table:trigram}
\end{table}

Note that our approach (the third column), i.e., our deltas and the trigram index, uses at most 1.5 of the memory spent on the storage of all commits (the second column). In two cases out of three, our data structure uses less memory, up to two times.

\paragraph{Trigram metrics}
For an illustration of trigram metrics, we chose the \textbf{xodus} repository.
The number of unique trigrams is $28 \cdot 10^3$ with the total count of $37 \cdot 10^6$. Since we treat the tabulation as the space, the most popular trigrams are: three spaces with $5 \cdot 10^6$ occurrences and the end of the line with two spaces with $7 \cdot 10^5$ of occurrences. The most popular trigrams composed solely of letters, are \texttt{ion}, \texttt{tio}, and \texttt{ing} with the number of occurrences $162$, $138$, and $113$ thousands, respectively.



\paragraph{Index request}
The most common request to our index takes a trigram and returns the newly created list of all files containing it (not the iterator).
Since as a key we store a pair of a trigram and a file~--- this operation is not trivial.
We measure a time spent on such request for trigrams with different frequencies.
For the most popular trigram \texttt{ion} with $1.6 \cdot 10^5$ occurrences the request takes $0.63$ milliseconds, for the trigram \texttt{bor} with $10^3$ occurrences the request is processed in $0.02$ milliseconds, while for one of the least popular trigram \texttt{abc} with $10$ occurrences the request takes $0.01$ milliseconds. So, we can conclude that the mostly used request is served really fast.

\begin{figure}[ht]
    \centering
    \includegraphics[width=0.99\linewidth]{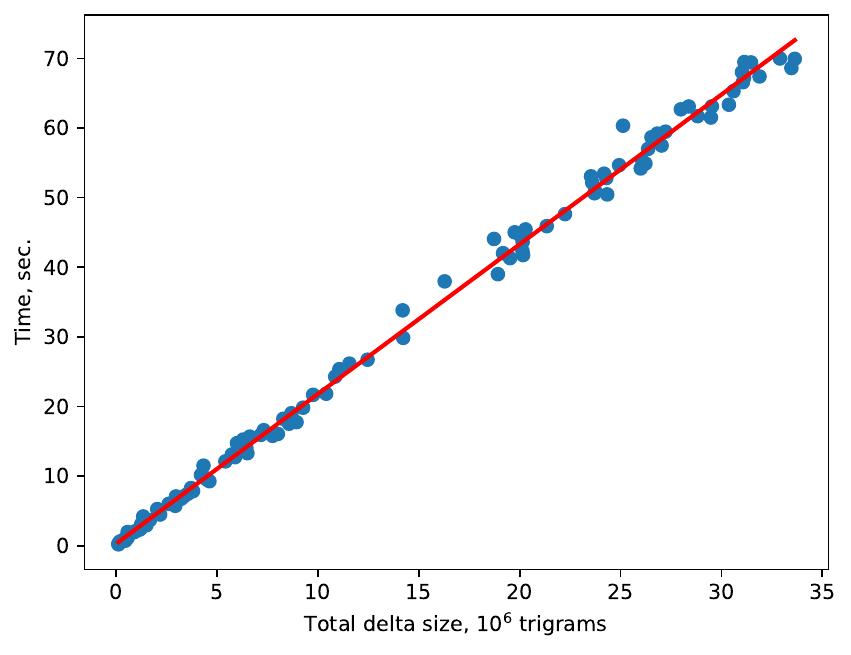}
    \caption{The time of the checkout operation as a function of trigrams number for \textbf{xodus} repository. For other repositories the dependence is exactly the same. It spends $2.15$ seconds for $10^6$ trigrams.}
    \label{fig:checkout}
\end{figure}

\paragraph{Checkout operation}
We also decided to measure the checkout time. We cannot run checkouts between all the pairs of revisions, i.e., the current and the checkouted versions, since their amount is just too large. Instead, we took just $10^3$ random pairs. We plot the dependency between the size of deltas in the number of trigrams between the versions and the time spent in Figure~\ref{fig:checkout}. The behavior is linear and is the same for all three chosen repositories with the time-usage of $2.2$ seconds per million trigrams. Even for the largest difference between the revisions (the latest commit and the starting point) our approach outperforms the construction of the index from scratch:
the time of checkout and building the trigram index from scratch for the given revision exceeded $70$ seconds.
In practice, the usual checkout is performed between two recent commits with small delta and takes an order of $0.1$ milliseconds, while the checkout to the month-old commit takes about one second.

\subsection{CamelHump index}
We measure the resources spent on building the CamelHump index for the symbols of the whole repository. The results appear in Table~\ref{table:camelhump}.

\begin{table}[ht]
\begin{tabular}{|l|c|c|c|c|}
\hline
            & \begin{tabular}{@{}c@{}}
                Number \\ of \\ commits
             \end{tabular} & \begin{tabular}{@{}c@{}}
                Total \\ memory
                \\
                of commits, \\ MBs                
             \end{tabular} & \begin{tabular}{@{}c@{}}
                Memory \\
                used, \\ MBs
             \end{tabular} & \begin{tabular}{@{}c@{}}
                Preprocess \\
                time, \\ sec            
             \end{tabular} \\\hline
\textbf{xodus} & 3180        & 253 & 73 & 350 \\ \hline
\begin{tabular}{@{}l@{}}
\textbf{commons-}\\
\textbf{lang}
\end{tabular} & 7653 & 765 & 98 & 484 \\ \hline
\textbf{maven} & 14275 & 453 & 145 & 718 \\ \hline
\end{tabular}
\caption{The initialization of the CamelHump index.
The number of commits and the total memory of commits are the characteristics of the repository. Memory used is the memory consumed by our data structure on the disk and the preprocessing time is the time for the initialization of the whole repository.}
\label{table:camelhump}
\end{table}

Note that, in comparison to the full-text trigram index, the memory used by our approach for the CamelHump search is significantly smaller than the memory used by all the commits as we store only the information for the symbols.

\paragraph{CamelHump trigrams metrics}
Again, exploring data from the \textbf{xodus} repository, we discovered $6.5 \cdot 10^3$ unique symbols. For CamelHump search, the number of unique trigrams total $10^4$, which is approximately $2.9$ times less than for the full-text search. This number is less due to the fact that we throw away the trigrams with white space symbols and upper case letters, which are stored only for full-text search. The total count of stored trigrams for all symbols sums up to $2.7 \cdot 10^5$, approximately $137$ times less than the full-text search count. This reduction is reasonable, considering only the unique words, significantly smaller than the entire text. 

\paragraph{Index request}
As for the trigram index, we measure the time spent on requesting the set of symbols for trigrams with different frequencies.
The most popular trigram in CamelHump index is \texttt{get} which appears $2470$ times. This number is one percent of the total number of trigrams. The time of the request for the trigram \texttt{get} is $2.78$ milliseconds.

Additionally, we examine the longest symbol in \textbf{xodus} repository, \texttt{getEntityIterableCacheHeavyIterablesCacheSize}.
The overall time spent on requests of all the trigrams in this symbol in the CamelHump index is $9.76$ milliseconds which is still sufficiently fast.


\section{Conclusion}
\label{sec:conclusion}

The proposed persistent trigram index enables support for various features in code review and seamless integration into modern cloud IDEs. The revision tree supports integration with version control systems and ensures on-disk persistence, aligning with our aim for zero-time startup. The algorithm ensures swift updates to the index during checkouts and commits. Additionally, we introduced an efficient approach to facilitate CamelHump search for symbols.

We conducted experimental testing of this solution across several open-source repositories. The results demonstrate that the requests in our data structure are fast and our approach efficiently utilizes the memory resources.

\begin{acks}
Vitaly Aksenov is partially sponsored by JetBrains.
\end{acks}

\bibliography{refs.bib}

\end{document}